\documentclass[prl,letterpaper,aps,twocolumn,floatfix]{revtex4}
\usepackage{graphicx}
\usepackage{amsmath}
\begin{document}
\title{Vortex-Peierls States in Optical Lattices}
\author{A.A. Burkov and Eugene Demler}
\affiliation{Department of Physics, Harvard University, Cambridge, 
Massachusetts 02138, USA}
\date{\today}
\begin{abstract}
We show that vortices, induced in cold atom superfluids in optical
lattices, may order in a novel vortex-Peierls ground state. In such
a state vortices do not form a simple lattice but arrange themselves in
clusters, within which the vortices are partially delocalized, tunneling
between classically degenerate configurations. We demonstrate that
this exotic quantum many-body state is selected by an
order-from-disorder mechanism for a special combination of the vortex
filling and lattice geometry that has a macroscopic number of
classically degenerate ground states.
\end{abstract}
\maketitle
The existence of quantized vortices is one of the most dramatic manifestations 
of the macroscopic wavefunction (``order parameter'') in superfluid (SF) Bose 
gases. 
Considerable theoretical \cite{theorvortex} and experimental \cite{expvortex} 
effort has thus been applied to the study of 
SF vortices in cold atoms systems.
Vortices are usually regarded as classical 
objects, that form regular lattices due to long-range repulsive interactions 
between them.
Intuitively, this point of view seems to be in accord with the fact that 
vortices are {\em topological objects}, i.e. the existence of an 
isolated vortex in the system can be established by observing phase winding 
infinitely far away from the vortex core.      
However, analogies with two-dimensional (2D) electronic systems in high 
magnetic fields, exhibiting the fractional quantum Hall (FQH) effect, 
suggest that under certain conditions vortex lattices may be melted by 
quantum fluctuations, and various strongly correlated vortex 
liquid states may thus emerge \cite{Gunn00}.

Optical lattices offer additional opportunities to explore the quantum 
mechanical behavior of vortices by allowing one to tune the strength 
of quantum fluctuations. 
Several approaches to stabilizing FHQ states in cold atoms systems using 
optical lattices have already been 
proposed \cite{Demler05,Mueller04,Zoller03}. 
More generally, of interest are situations in which 
vortices may behave as strongly interacting {\em quantum particles}, 
moving in a periodic optical lattice potential.
Such a possibility appears to be rather counterintuitive, 
since naively we think of vortices as macroscopic objects.
Nevertheless, the common view nowadays is that vortices in 2D systems can 
be considered as quantum particles with a finite mass \cite{Fazio01}.  
It is then interesting to find experimentally observable phenomena in cold 
atoms systems in which this quantum nature of vortices is manifested. 
In particular, manifestations of the quantum mechanical behavior of 
vortices may be most dramatic in situations when long-range intervortex 
interactions are {\em frustrated}, which strongly enhances the effect 
of quantum fluctuations. Such a frustration in optical 
lattice systems may be engineered by choosing the appropriate 
combination of the optical lattice geometry and the vortex density.   
We should point out that some classical commensuration effects between 
vortex lattices and the underlying optical lattice pinning potential have 
already been studied, both 
theoretically \cite{Wu04} and experimentally 
\cite{Cornell06}, but the possible quantum effects have not been previously 
considered.

In this Letter we discuss an example of such a quantum mechanical 
behavior of SF vortices in an optical lattice. 
We show that for a particular combination of the 
optical lattice geometry and the vortex filling, for which classical 
vortex configurations are strongly frustrated, a {\em vortex-Peierls} (VP)
state is realized.
By VP state we mean a vortex lattice, in which vortices are not localized 
at the maxima of the optical lattice potential, but are instead partially 
delocalized, resonating quantum mechanically between degenerate  
pinned configurations.
Peierls ordering has been extensively studied, most recently in the context 
of quantum magnetism of localized spin systems (see Ref.\cite{Sachdev} for 
review). Here we demonstrate that such ordering may under certain conditions 
occur for vortices in optical lattice superfluids.
   
\begin{figure}[t]
\includegraphics[width=4cm]{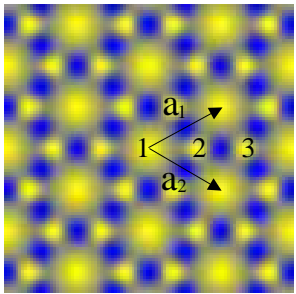}
\caption{(Color online).
Optical dice lattice created by superimposing three laser field potentials 
$I_1({\bf r})$, $I_2({\bf r})$ and $I_3({\bf r})$ (see text).
${\bf a}_1 = \frac{3}{2} \hat x - \frac{\sqrt{3}}{2} \hat y$ and 
${\bf a}_2 = \frac{3}{2} \hat x + \frac{\sqrt{3}}{2} \hat y$ are the basis 
directions. Lighter areas correspond to potential minima.
Inequivalent sites in the three-site unit cell of the dice lattice are 
labeled as 1,2 and 3.} 
\label{fig:dice}
\vspace{-0.5cm}
\end{figure}

We consider a SF system of cold bosonic atoms, loaded in an optical 
periodic potential with the {\em dice lattice} geometry, 
shown in Fig.\ref{fig:dice}. The fascinating features of the quantum 
mechanics of particles on the dice lattice in a perpendicular magnetic field 
were first pointed out by Vidal {\it et al.} \cite{Vidal98} and extensively 
studied in a number of subsequent works (see e.g.\cite{Fazio05} and 
references therein).     
Atoms in an optical lattice can be described by the Bose-Hubbard 
model \cite{Zoller98}, in which bosons are assumed to tunnel between 
nearest-neighbor sites of the lattice and interact when they are on the 
same site.   
We will assume that the average number of bosonic atoms per site of the 
dice lattice $\bar n$ is an integer. 
In this case increasing the on-site interaction energy $U$ relative to the 
hopping amplitude $t$ until $U/\bar n t \sim 1$ will induce a 
SF-Mott Insulator (MI)
transition \cite{Greiner02} when $\bar n$ bosons will be localized 
on each site to minimize the interaction energy. 

To induce vortices in the SF we add an effective perpendicular 
``magnetic field''. 
We will comment on particular methods, that could be used to create 
such an effective field, at the end of the paper. 
We will focus on a specific value of the flux per plaquette of the dice 
lattice $2 \pi f = 2 \pi/3$, where $f=1/3$ has the physical meaning of the 
vortex filling, i.e. the average number of vortices per plaquette. 
Centers of dice plaquettes can be associated with sites of the dual kagom\'e
lattice, which correspond to the maxima of the optical lattice potential.  
It is then convenient to assume that the vortex cores are located 
on the kagom\'e lattice sites.   
Since vortices interact via a long-range repulsive potential,
they will try to arrange themselves in patterns on the kagom\'e lattice, that 
maximize the distance between each vortex and its neighbors.
As shown by Korshunov \cite{Korshunov05}, at filling factor $f=1/3$ the set 
of vortex configurations that minimize the classical interaction energy 
between the vortices consists of all  
states, where every triangular plaquette of the kagom\'e lattice is 
occupied by exactly one vortex.  
The number of such configurations grows exponentially with the system size 
and the classical ground state of vortices has only algebraic 
order at zero temperature, but no true long-range order \cite{Korshunov05}.
 
Nevertheless, below we will demonstrate that strong quantum fluctuations near 
the SF-MI transition lift the classical degeneracy and select a 
vortex state with a true long-range order.  
This state has, however, a manifestly quantum mechanical nature, 
in that the vortices are not simply localized on sites of the 
kagom\'e lattice, but are instead partially delocalized over plaquettes, 
thus exhibiting what we call a VP ordering. 

To approach the problem analytically, we will take advantage of the presumed
proximity of our system to the SF-MI critical point.
Since the bosons are at an integer filling, the standard 
Landau-Ginzburg (LG) theory can be applied not too far away from the critical 
point.
The imaginary time LG functional of the SF order 
parameter fields is given by \cite{Altman02}:
\begin{eqnarray}
\label{eq:3}
S&=&\int_0^{\beta} d \tau \left[ - J \sum_{i\mu} (\Phi^*_i 
\Phi^{\vphantom *}_{i+\mu} e^{-i A_{i\mu}} + c.c.) \right. \nonumber \\ 
&+&\sum_i(|\partial_{\tau} \Phi_i |^2 
 + \left.r |\Phi_i|^2 + u_4 |\Phi_i|^4 + u_6 |\Phi_i|^6)\right].
\end{eqnarray}
Here $\Phi_i$ is the local SF order parameter,
$\mu$ denote the nearest-neighbor vectors of the dice lattice, 
$J \sim U \bar n t$, $u_4, u_6 > 0$ and $r$ tunes the system across the 
SF-MI transition. 
To find the ordering patterns near the transition, we need to diagonalize 
the first term in Eq.(\ref{eq:3}) \cite{Berker84}.
Choosing Landau gauge for the vector potential
$A_{i\mu} = 2 \pi f i_x (1 - \delta_{\mu x})$
one obtains the following dispersion for the lowest Hofstadter band:
$\epsilon({\bf k}) = - J \left[6 + \cos(k_1) + \cos(k_2) 
- 2 \cos(k_1 - k_2)
+ \sqrt{3} \sin(k_1) \right. \\ + \left. \sqrt{3} \sin(k_2) \right]^{1/2}$,
where ${\bf k} = k_1 {\bf b}_1 + k_2 {\bf b}_2$ and $
{\bf b}_{1,2} = \frac{1}{3} \hat x \pm \frac{1}{\sqrt{3}} \hat y$
are the reciprocal lattice vectors of the dice lattice. 
The boson dispersion has two minima inside the 
first Brillouin zone of the dice lattice, at wavevectors 
${\bf k}_0 = (0, 2\pi /3)$ and ${\bf k}_1 = (2 \pi/3, 0)$.
The corresponding eigenvectors are given by 
$v^0 = (1, 1, 0)$ and $v^1 = (e^{-2 \pi i /3}, 0 ,1)$.
We can then write the lattice order parameter fields $\Phi_i$ as linear 
combinations of these two low energy modes
$ \Phi_{\sigma}({\bf r}_i) = \sum_{\ell = 0,1} \varphi_{\ell} v^{\ell}_{\sigma}
e^{i {\bf k}_{\ell} \cdot {\bf r}_i}$,
where the index $i$ labels the unit cells of the dice lattice, 
$\sigma = 1,2,3$ labels sites within 
each unit cell and $\varphi_{\ell}$ are the fields, corresponding to the low
energy boson modes.
To obtain the LG action in terms of
the fields $\varphi_{\ell}$ in its most general form we need to know how 
these fields transform under the symmetry operations of the dice lattice. 
The relevant operations are the elementary translations along 
the basis directions $T_1$ and $T_2$, rotations by $\pi/3$ around the six-fold 
coordinated sites $R_{\pi/3}$ and reflections with respect to the $x$- and
$y$-axes $I_{x,y}$.
These transformations are given by:
$T_1 : \varphi_{\ell} \rightarrow \varphi_{\ell} e^{-2 \pi i \ell /3 },\,\,
T_2 : \varphi_{\ell} \rightarrow \varphi_{\ell} e^{2 \pi i (\ell-1)/3},\,\, 
R_{\pi/3} : \varphi_{\ell} \rightarrow \varphi_{\ell + 1} 
e^{-2 \pi i (\ell+1)/3}, \,\, I_x : \varphi_{\ell} \rightarrow 
\varphi_{\ell}^* e^{-2 \pi i \ell/3}, \,\,
I_y : \varphi_{\ell} \rightarrow \varphi_{\ell+1}^* e^{2 \pi i/3}$,
where the subscripts of the fields are taken modulo 2. 
Using these transformations, the most general form of the imaginary time 
LG action is found to be \cite{Kim06}:
\begin{eqnarray}
\label{eq:11}
S&=&\int_0^{\beta} d \tau \int d {\bf r}
\left\{\sum_{\ell}[|\partial_{\tau} \varphi_{\ell}|^2 + 
c^2 |\partial_{\mu} \varphi_{\ell}|^2 + r |\varphi_{\ell}|^2]\right. 
\nonumber \\
&+&\left.u_4 (\sum_{\ell} |\varphi_{\ell}|^2)^2 + 
v |\varphi_0|^2 |\varphi_1|^2 \right.\nonumber \\
&+& \left.u_6 (\sum_{\ell} |\varphi_{\ell}|^2)^3 + 
w [(\varphi_0^* \varphi_1^{\vphantom *})^3 + c.c.] \right\}.
\end{eqnarray}
\begin{figure}[t]
\includegraphics[width=7cm]{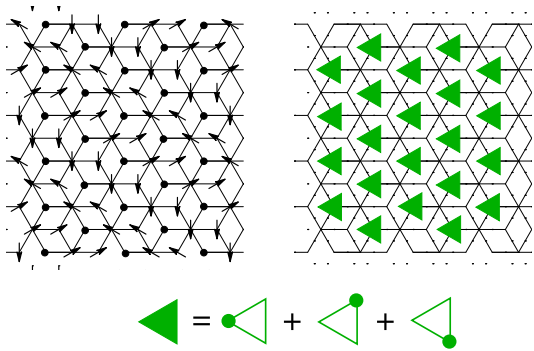}
\caption{(Color online). (Left) Order parameter configuration corresponding to 
the $v > 0$ state. Direction of each arrow represents the phase of the 
corresponding order parameter field, while length represents the magnitude. 
Expectation values of the order parameter vanish on all sites of 
type 3 (shown by dots).
(Right) Corresponding vortex configuration. Shaded triangles contain a vortex, 
uniformly delocalized over the three sites of each triangle. Note that 
the supercurrents vanish on every bond of the dice lattice.} 
\label{fig:vpos}
\end{figure}
\begin{figure}[t]
\includegraphics[width=7cm]{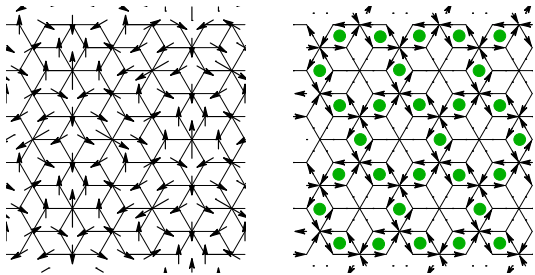}
\caption{(Color online). (Left) Order parameter configuration corresponding to 
the $v,w < 0$ state.
(Right) Corresponding supercurrent configuration. 
Plaquettes, containing vortices (shown by circles), are the ones that have 
supercurrents circulating around them in the counterclockwise direction.} 
\label{fig:vnegwneg}
\vspace{-0.5cm}
\end{figure}
\begin{figure}[t]
\includegraphics[width=7cm]{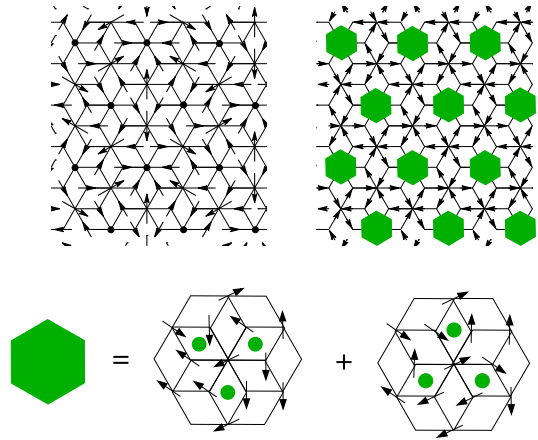}
\caption{(Color online). (Left) Order parameter configuration corresponding to 
the $v < 0, w > 0$ state. 
Expectation value of the order parameter vanishes on some of the 
six-fold coordinated sites (shown by dots). 
(Right) Corresponding supercurrent configuration. 
Note that none of the plaquettes contain a full vortex. Instead, 
vortices are bound into triplets populating the groups of six plaquettes, 
adjacent to six-fold coordinated sites with zero order parameter expectation 
value.
Below, SF order parameter phase snapshots, 
corresponding to the two resonating vortex triplet configurations are shown. 
} 
\label{fig:vnegwpos}
\vspace{-0.5cm}
\end{figure} 
In the mean-field approximation, which we expect to be accurate 
for the effective classical 2+1-dimensional system, described by 
Eq.(\ref{eq:11}), three different SF 
phases are possible, depending on the signs of the $v$ and $w$ couplings. 
\begin{enumerate}
\item $v > 0$: either $|\varphi_0| \ne 0 $ or $|\varphi_1| \ne 0$.
\item $v < 0, w < 0$: $|\varphi_0| = |\varphi_1| \ne 0$. 
The relative phase 
$\theta = \arg[\varphi_0^* \varphi_1^{\vphantom *}]$  
is determined by the last term in Eq.(\ref{eq:11}) and is given by
$\theta = 2 \pi n/3,\,\, n=0,1,2$.
\item $v < 0, w > 0$: $|\varphi_0| = |\varphi_1| \ne 0$.
The relative phase is given by $\theta = (2 n+1) \pi /3,\,\, n=0,1,2$.
\end{enumerate}
To reinterpret the states we have found in the vortex language, it is 
convenient to calculate gauge-invariant supercurrents on each bond, which we 
define as
$J_{i\mu} \sim \textrm{Im} \left(\Phi^*_i \Phi^{\vphantom *}_{i+\mu}
e^{-i A_{i\mu}}\right)$.
In the $v > 0$ state we find that supercurrents vanish on every bond. 
This fact, combined with the picture of this state in terms of the SF 
order parameter, leads to the vortex configuration shown in the right panel of 
Fig.\ref{fig:vpos}.  
Since this state contains configurations, in which two vortices 
are located on nearest-neighbor sites, it can not be the true ground 
state of the vortices. 

Calculating supercurrents in the second candidate ground state, realized 
when $v,w < 0$, we obtain the configuration shown in the right panel of 
Fig.\ref{fig:vnegwneg}.
Vortices in this case are localized on the dice lattice plaquettes, which 
have supercurrents circulating around them in the counterclockwise direction.
This configuration is a member of the classical ground state manifold, 
since none of the vortices have nearest neighbors. 

Finally, the supercurrent pattern in the state, realized when $v < 0$ and 
$w > 0$, is shown in the right panel of Fig.\ref{fig:vnegwpos}.
One can see that none of the dice plaquettes in this case 
have a full vortex localized in it.
Instead the vortices appear to be bound in partially delocalized triplets, 
populating the six 
dice lattice plaquettes, adjacent to the six-fold coordinated sites 
with zero order parameter expectation values. 
The fact that the order parameter vanishes on these sites 
means that vortices are moving around such sites, locally destroying phase 
coherence.
It also means that the boson number does not fluctuate on such sites, 
i.e. the bosons on these sites are in the Mott phase.
The only vortex state that is consistent with this picture, and also does not
violate the constraint of no-nearest-neighbor vortices is the one 
in which the vortex triplets resonate between two degenerate configurations 
(corresponding instantaneous order parameter phase configurations 
are explicitly shown in Fig.\ref{fig:vnegwpos}).
This state is clearly more energetically favorable than the state in 
Fig.\ref{fig:vnegwneg},
since it does not violate the no-nearest-neighbor constraint but also 
allows the vortices to gain kinetic energy by partially delocalizing over 
hexagonal plaquettes of the dual kagom\'e lattice.  
We can estimate the energy gain in this state due to the vortex delocalization,
compared to the $w < 0$ state, as follows.
Interaction energy of the vortices (energy of the supercurrents) is of 
order $\bar n t$ per lattice site. Vortex kinetic energy (energy gain from 
phase fluctuations) is of order $U$ per site.
Therefore, the energy gain in the $w > 0$ state, compared to the $w < 0$ 
state, is of order $\bar n t (U/\bar n t)^3$, which is not small 
when $U \sim \bar n t$.
Thus, we find that the state in Fig.\ref{fig:vnegwpos} is the ground state 
of the vortices in our problem. 
Since the vortices are not localized on sites of the kagom\'e lattice, but are 
partially delocalized over plaquettes, we call this state a VP state. 

Let us now discuss how to observe the VP state experimentally. 
Creating an optical lattice with a dice geometry experimentally is more 
difficult than most other 2D lattices, but fortunately still possible with 
the current technology. We propose the following procedure 
(see Ref.\cite{Fazio05} for an alternative proposal). 
One first creates a kagom\'e lattice, which is done using the laser field 
potential proposed in Ref.\cite{Lewenstein04}:
$I_1({\bf r}) = \sum_{i=1}^3 \left[\cos\left({\bf k}_i \cdot {\bf r} + 
\frac{\pi \sigma_i}{2}\right) 
 + 2 \cos\left(\frac{1}{3} {\bf k}_i \cdot 
{\bf r} + \frac{3 \pi \sigma_i}{2}\right)\right]^2$,
where ${\bf k}_1 = (\pi, \sqrt{3} \pi)$, ${\bf k}_2 = (\pi, -\sqrt{3} \pi)$, 
${\bf k}_3 = (-2 \pi, 0)$, $\sigma_1 = \sigma_3 = 1$ and 
$\sigma_2 = -1$.
Here the {\em maxima} of the potential correspond to the kagom\'e lattice 
sites.
To create a perfect dice lattice we superimpose two additional laser 
potentials, that have a triangular lattice geometry: 
$I_2({\bf r}) =  4 \sum_{i=1}^3 \cos^2\left(\frac{1}{3} {\bf k}_i \cdot 
{\bf r}\right)$, and 
$I_3({\bf r}) =  - 4 \sum_{i=1}^3 \cos^2\left({\bf g}_i \cdot 
{\bf r}\right)$,
where ${\bf g}_1 = (\pi,\pi/\sqrt{3})$, ${\bf g}_2 = (0, -2 \pi/\sqrt{3})$, 
and ${\bf g}_3 = (-\pi, \pi/\sqrt{3})$.
Here $I_2({\bf r})$ has maxima at the six-fold coordinated sites of the 
dice lattice, while $I_3({\bf r})$ has minima at both three-fold and six-fold 
coordinated sites. 
The superposition of $I_1({\bf r}), I_2({\bf r})$ and $I_3({\bf r})$ creates 
a perfect dice lattice, in which all potential wells have equal depth.
Effective magnetic flux in this setup can be created by the rotating mask 
method \cite{Cornell06}, which generates a rotating optical lattice potential.

Alternatively, the proposal of Ref.\cite{Demler05} can be used to create 
the effective perpendicular field. 
One uses the combination of a time-dependent quadrupolar potential 
$V(t) = V_{qp} \sin(\omega t) xy$ and a temporal modulation of the tunneling
amplitudes for different nearest-neighbor directions in the dice lattice. 
Modulation of the tunneling amplitudes can be achieved in our case by 
varying the strengths of the three components of the $I_3({\bf r})$ potential. 
To achieve the flux of $2\pi/3$ per plaquette, the parameters of the 
quadrupolar potential have to be chosen such that 
$V_{qp}/\hbar \omega = \pi/2 \sqrt{3}$.
In this setup, 
time-of-flight interference imaging can be used to study the periodic 
structure of the vortex lattice. Lattice periodicity, however, is not enough 
to distinguish 
the VP state we found from the state in Fig.\ref{fig:vnegwneg},
since they have identical reciprocal lattice vectors. 
VP ordering can be detected 
by analyzing nontrivial {\em noise correlations} \cite{Altman04}, that 
will be present in the time-of-flight image.
Namely, in addition to the sharp peaks in the density distribution (i.e. first 
order correlation function), corresponding to the reciprocal 
lattice vectors ${\bf g}_i/3$ of the votex lattice, one should observe 
strong incoherent background. Such a background would be absent in a regular 
superfluid, in which all vortices are strictly localized.  
This is due to the fact that the vortex motion locally destroys phase 
coherence on some sites of the optical lattice. 
In the case of the VP 
state, described above, the incoherent background contains correlations at 
the same wavevectors ${\bf g}_i/3$, which may be found by measuring the 
density autocorrelation function.  
If the perpendicular field is created by the rotating mask technique,
one can use interference between two identical co-rotating condensates
\cite{Altman05} to image the vortex lattice configurations. 

In conclusion, we have proposed that vortices in optical lattice 
SF may exist in VP ground states, which are 
direct analogs of valence-bond-solid states of interacting bosons. 
In particular, we have demonstrated that in the case of a dice optical 
lattice with vortex filling of $1/3$ per plaquette, the ground state of the 
vortices is a plaquette VP state, in which vortices bind into 
triplets, that resonate between two degenerate configurations on plaquettes
of the dual kagom\'e lattice. Such unconventional vortex ordering is a result 
of an order-by-disorder phenomenon, where extensive degeneracy of frustrated
classical vortex configurations is lifted by quantum fluctuations. 

\acknowledgments{We would like to thank E.~Altman, M.~Greiner, B.I.~Halperin,
M.D.~Lukin, R.G.~Melko and V.~Schweikhard for useful discussions. 
Financial support was provided by the National Science Foundation 
under grants DMR02-33773 and DMR01-32874.}

\end{document}